\documentclass[manuscript]{aastex}
\newcommand{\age}{t_{\rm age}}
\newcommand{\tn}{\tilde{N}}

\usepackage{amsmath, amssymb}

\slugcomment{Not to appear in Nonlearned J., 45.}

\shorttitle{Magnetar Wind Nebulae}
\shortauthors{S. J. Tanaka}

\begin{document}


\title{A Broadband Emission Model of Magnetar Wind Nebulae}


\author{Shuta J. Tanaka}
\affil{
Department of Physics, Faculty of Science and Engineering, Konan University, 8-9-1 Okamoto, Kobe, Hyogo 658-8501, Japan
}
\affil{
Institute for Cosmic Ray Research, The University of Tokyo, 5-1-5 Kashiwa-no-ha, Kashiwa City, Chiba, 277-8582, Japan
}
%
%




\begin{abstract}
Angular momentum loss by the plasma wind is considered as a universal feature of isolated neutron stars including magnetars.
The wind nebulae powered by magnetars allow us to compare the wind properties and the spin-evolution of magnetars with those of rotation-powered pulsars (RPPs).
In this paper, we construct a broadband emission model of magnetar wind nebulae (MWNe).
The model is similar to past studies of young pulsar wind nebulae (PWNe) around RPPs, but is modified for the application to MWNe that have far less observational information than the young PWNe.
We apply the model to the MWN around the youngest ($\sim$ 1kyr) magnetar 1E 1547.0-5408 that has the largest spin-down power $L_{\rm spin}$ among all the magnetars.
However, the MWN is faint because of low $L_{\rm spin}$ of 1E 1547.0-5408 compared with the young RPPs.
Since most of parameters are not well constrained only by an X-ray flux upper limit of the MWN, we adopt the model parameters from young PWN Kes 75 around PSR J1846-0258 that is a peculiar RPP showing magnetar-like behaviors.
The model predicts $\gamma$-ray flux that will be detected in a future TeV $\gamma$-ray observation by {\it CTA}.
The MWN spectrum does not allow us to test hypothesis that 1E 1547.0-5408 had milliseconds period at its birth because the particles injected early phase of evolution are suffered from severe adiabatic and synchrotron losses.
Further both observational and theoretical studies of the wind nebulae around magnetars are required to constrain the wind and spin-down properties of magnetars.
\end{abstract}





\section{INTRODUCTION}\label{sec:intro}

Rotation-powered pulsars (RPPs) convert most of their rotational energy into pulsar winds and create pulsar wind nebulae (PWNe) around them \citep[e.g.,][]{Pacini&Salvati73, Rees&Gunn74}.
More than seventy pulsar wind nebulae (PWNe) have been detected so far and most of them are around energetic RPPs whose spin-down powers $L_{\rm spin}$ are $\gtrsim 10^{35} {\rm erg~s^{-1}}$ \citep[c.f.,][]{Kargaltsev+13}.
However, some less energetic isolated pulsars ($L_{\rm spin} \gtrsim 10^{33} {\rm erg~s^{-1}}$) are also associated with bow shock structures, called bow shock PWNe, as a consequence of the wind from the pulsars \citep[e.g.,][]{Kargaltsev+15}.
It is considered that not only the rotational energy loss by pulsar winds but also the formation of PWNe are the universal features of isolated pulsars.

Magnetars are a class of isolated pulsars and have the inferred surface magnetic field strength above the quantum critical field $B_{\rm QED} \equiv m^2_{\rm e} c^3 / e \hbar \approx 4.4 \times 10^{13} {\rm G}$ \citep[e.g.,][]{Mereghetti08}.
Their large persistent X-ray luminosity exceeding the spin-down power $L_{\rm spin}$ and bursting activities in soft $\gamma$-rays indicate that they are magnetically-powered objects \citep[][]{Thompson&Duncan95, Thompson&Duncan96}.
On the other hand, some magnetars are also pulsating in radio as RPPs \citep[e.g.,][]{Camilo+06}.
This is the evidence that plenty of electron-positron pairs are produced in the magnetar magnetosphere \citep[][]{Medin&Lai10, Beloborodov13} against the photon splitting process in the intense magnetic field \citep[][]{Baring&Harding98}.
In addition, it is discussed that the pulsed radio emission from magnetars is rotation-powered \citep[][]{Rea+12, Szary+15}.
Besides magnetic energy responsible for bright X-rays and bursting activities, magnetars also release their rotational energy $L_{\rm spin}$ as the angular momentum extraction by the magnetar wind.

Detection of magnetar wind nebulae (MWNe) is one of the best way to confirm the presence of the magnetar wind.
Although some report detections of MWN-like extended emission around AXP 1E 1547.0-5408 \citep[][]{Vink&Bamba09} and Swift J1834.9-0846 \citep[][]{Younes+12}, others claim that the extended emission is consistent with dust-scattering halos of their past magnetar activities \citep[][]{Olausen+11, Esposito+13}.
However, we have obtained upper limits on MWN emissions, which are still important to give a constraint on the presence of MWNe.
Combining deep observations of magnetars with spectral studies of PWNe, we would be able to constrain pair-production multiplicity $\kappa$ \citep[c.f.,][]{deJager07, Bucciantini+11} and magnetization $\sigma$ \citep[c.f.,][]{Martin+14} of the magnetar winds, and also the spin-down evolution of the central magnetars \citep[c.f.,][]{deJager08}.
Especially, it is interesting to explore differences of them from RPPs, for example, $\sigma$- and $\kappa$-problems are known for young RPPs \citep[c.f.,][]{Kirk+09, Arons12, tt13a}.

Studies of MWNe have another interesting aspect.
Long-lasting activities of gamma-ray bursts ($< 10^5$s) are often interpreted as due to the spin-down activity of a magnetar born with milliseconds period \citep[e.g.,][]{Zhang&Meszaros01} although there are some other models \citep[c.f.,][]{Kisaka&Ioka15}.
In addition, millisecond magnetars are also proposed as the engines of gamma-ray bursts \citep[e.g.,][]{Usov92} and luminous supernovae \citep[c.f.,][]{Kashiyama+15}.
On the other hand, population synthesis studies of magnetars show the difficulty to be compatible with the Galactic magnetar population with such millisecond magnetars although population studies are not easy to account for initial spin periods of less than 10 ms \citep[][]{Rea+15}.
Recently, very early phase ($< 10^6$ s) of the system of an embryonic MWN and ejecta of an explosion around a newly-born millisecond magnetar have been studied to find the evidence of millisecond magnetars \citep[e.g.,][]{Metzger+14, Murase+15}.
Here, we study the spectrum of MWNe at an age of $\sim$ kyr, where it may hold some signatures of millisecond magnetars.

In Section \ref{sec:Model}, we describe our model of the MWN spectral evolution.
We consider the persistent magnetar outflow as well as the pulsar wind and the model is similar to past studies of young PWNe around RPPs \citep[][hereafter TT10, TT11 and TT13b, respectively]{tt10, tt11, tt13b}.
In Section \ref{sec:1E1547}, we apply the model to the MWN around 1E 1547.0-5408.
1E 1547.0-5408 is the only object that we find the flux upper limit on the published literature and is the most promising object to detect MWNe among all the magnetars because of its large $L_{\rm spin}$ and small distance.
In Section \ref{sec:dis&cons}, the results are discussed in view of the connection between RPPs and magnetars.
Especially, we compare the MWN around 1E 1547.0-5408 with PWN Kes 75 around PSR J1846-0258 that is a RPP with large surface magnetic field and has experienced magnetar-like bursts in 2006 \citep[][]{Gavriil+08}.
We also discuss the existence of millisecond magnetars and conclude this paper.

\section{Model}\label{sec:Model}

Here, based on the presumption that magnetars also create wind nebulae like young RPPs, a one-zone spectral model of MWNe is introduced.
The energy distribution of accelerated electrons and positrons inside MWNe $N(\gamma, t)$ is found from the continuity equation,
\begin{eqnarray}\label{eq:BoltzmannEquation}
	\frac{\partial}{\partial t}      N(\gamma, t)
	+
	\frac{\partial}{\partial \gamma} \left( \dot{\gamma}(\gamma, t) N(\gamma, t) \right)
	& = &
	Q_{\rm inj}(\gamma, t),
\end{eqnarray}
where $\gamma$ is the Lorentz factor of the relativistic electrons and positrons, $Q_{\rm inj}(\gamma, t)$ is the injection from the central magnetar and $\dot{\gamma}(\gamma, t)$ includes adiabatic $\dot{\gamma}_{\rm ad}(\gamma,t)$, synchrotron $\dot{\gamma}_{\rm syn}(\gamma,t)$, and inverse Compton scattering $\dot{\gamma}_{\rm IC}(\gamma)$ coolings.
Although most of concepts are shared with ours and other past studies of PWN spectra \citep[][]{Gelfand+09, Bucciantini+11, Martin+14, Torres+14}, we introduce some modifications for the application to MWNe that have much less observational information than young PWNe.

The radiation processes are synchrotron radiation and the inverse Compton scattering off the interstellar radiation field (IC/ISRF) and we ignore the synchrotron self-Compton (SSC) process because SSC is always sub-dominant to IC/ISRF for young PWNe other than the Crab Nebula \citep[c.f.,][]{Torres+13}.
The ISRF has three components: the cosmic microwave background radiation (CMB), dust infrared (IR) and optical starlights (OPT).
The CMB has the blackbody spectrum of the temperature $T_{\rm CMB} = 2.7$ K and the others are assumed to have the modified blackbody spectra which are characterized by energy densities $u_{\rm ISRF}$ and temperatures $T_{\rm ISRF}$.
Below, we use $(u_{\rm IR}, T_{\rm IR}) = (1.0 {\rm eV~cm^{-3}}, 40 {\rm K})$ and $(u_{\rm OPT}, T_{\rm OPT}) = (2.0 {\rm eV~cm^{-3}}, 4000 {\rm K})$ for the application to 1E 1547.0-5408.

\subsection{Spin-down evolution}\label{sec:SpinDownEvolution}

The spin-down of pulsars is customarily described as the differential equation $\dot{\Omega} = - k \Omega^n$, where $\Omega$ and $\dot{\Omega}$ are the current angular frequency and its derivative, respectively.
We apply the same equation to magnetars.
To specify the spin-down behavior $\Omega(t)$ from the differential equation, we need one initial condition and two constants $n$ and $k$.
We take the corresponding three quantities as a braking index $n$, an initial period $P_0 = 2 \pi / \Omega_0$ and an initial dipole-magnetic field $B_0 \equiv 3.2 \times 10^{19} {\rm G} \sqrt{P_0 \dot{P}_0}$.
Assuming that the moment of inertia of a magnetar is $I = 10^{45} {\rm g~cm^2}$, evolution of the spin-down power $L_{\rm spin}(t) = I \Omega(t) \dot{\Omega}(t)$ is expressed as
\begin{eqnarray}\label{eq:Lspin}
	L_{\rm spin}(t)
	& = &
	L_0 \left( 1 + \frac{t}{t_0} \right)^{-\frac{n+1}{n-1}},
\end{eqnarray}
where an initial spin-down power $L_0 = I \Omega_0 \dot{\Omega}_0$ and an initial spin-down time $t_0 = \Omega_0 / (1 - n) \dot{\Omega}_0$ relate with $P_0$ and $B_0$ as
\begin{eqnarray}
	L_0
	& = &
	3.9 \times 10^{43} {\rm erg~s^{-1}}
	\left( \frac{B_0}{10^{14} {\rm G} } \right)^2
	\left( \frac{P_0}{10      {\rm ms}} \right)^{-4}, \label{eq:L0} \\
	t_0
	& = &
	\frac{3.2}{(n - 1)} \times 10^{-4} {\rm kyr}
	\left( \frac{B_0}{10^{14} {\rm G} } \right)^{-2}
	\left( \frac{P_0}{10      {\rm ms}} \right)^2, \label{eq:t0}
\end{eqnarray}
respectively.
We also obtain the simple relation between the initial spin-down time $t_0$, the age $\age$ and the characteristic age $t_{\rm c}$ as
\begin{eqnarray}\label{eq:CharacteristicAge}
	t_{\rm c}
	& = &
	\frac{n - 1}{2} (\age + t_0),
\end{eqnarray}
where $t_{\rm c} \equiv P / 2 \dot{P}$ is obtained from $P$ and $\dot{P}$ at an age of $\age$.

For magnetars of $P_0 \lesssim 10$ ms and $B_0 \gtrsim 10^{14}$ G, because observed magnetars have an age of $\age \gtrsim$ kyr $\gg t_0$, Equations (\ref{eq:Lspin}) and (\ref{eq:CharacteristicAge}) are simplified as
\begin{eqnarray}\label{eq:LspinAgeApprox}
	L_{\rm spin}(t)
	& \approx &
	L \left( \frac{t}{\age} \right)^{-\frac{n+1}{n-1}},
	~~
	\age
	\approx
	\frac{2 t_{\rm c}}{n - 1}.
\end{eqnarray}
The present spin-down power $L = 4 \pi^2 I \dot{P} / P^3$ and the characteristic age $t_{\rm c} = P / 2 \dot{P}$ are obtained from observed values of $P$ and $\dot{P}$.
Only the braking index $n$ is the parameter of the magnetar spin-down within this approximation.
We take $n = 3$ ($\age \approx t_{\rm c}$) as a fiducial value, although the observed values of some pulsars have variation \citep[e.g., Table 1 of][]{Espinoza+11}.

\subsection{Expansion \& Magnetic Field Evolution}\label{sec:MagneticField}

We assume that a MWN is an expanding uniform sphere whose radius is expressed as
\begin{eqnarray}\label{eq:ExpansionEvolution}
	R_{\rm MWN}(t)
	& \approx &
	R \left( \frac{t}{\age} \right)^{\alpha_{\rm R}},
\end{eqnarray}
where $R$ is the present radius of the MWN and is estimated from observations.
Equation (\ref{eq:ExpansionEvolution}) with the index $\alpha_{\rm R} \gtrsim 1$ expresses an early-phase of expansion evolution \citep[c.f.,][]{Reynolds&Chevalier84, vanderSwaluw+01, Chevalier05, Gelfand+09}.
As discussed in Appendix \ref{app} (see Figure \ref{fig:AlphaBRDependence}), our model spectrum is insensitive to $\alpha_{\rm R}$.
We adopt $\alpha_{\rm R} = 1.0$ for the application to 1E 1547.0-5408.

Mean magnetic field strength inside a MWN should evolve with time because of magnetic energy injection from central magnetars and of expansion of the MWN.
For simplicity, we also adopt a power-law dependence on time for magnetic field evolution, i.e.,
\begin{eqnarray}\label{eq:BfieldEvolution}
	B_{\rm MWN}(t)
	& \approx &
	B \left( \frac{t}{\age} \right)^{\alpha_{\rm B}},
\end{eqnarray}
where $B$ is the current magnetic field strength of the MWN and is a parameter.
The current magnetic field strength ranges $3 \mu {\rm G} \lesssim B \lesssim 80 \mu {\rm G}$ for young PWNe \citep[c.f., TT13b;][]{Torres+14}.
Although the index $\alpha_{\rm B} \sim -2$ is different between models \citep[c.f., TT10;][]{Torres+14}, our model spectrum is insensitive to $\alpha_{\rm B}$ (see the discussion in Appendix \ref{app} and Figure \ref{fig:AlphaBRDependence}).
We will take $\alpha_{\rm B} = -1.5$ for the application to 1E 1547.0-5408.

In the past studies, the magnetic energy fraction $\eta$ has been used to determine the magnetic field strength $B$ of young PWNe, where the magnetic power of the central pulsars is expressed as $\eta L_{\rm spin}$.
TT11 and \citet{Martin+14} studied the value of $\eta$ for each object because $\eta$ is closely related with the wind magnetization parameter $\sigma$ \citep[the ratio of Poynting flux to particle energy flux of the pulsar wind, c.f.,][]{Kennel&Coroniti84}.
On the other hand, in the current model, we estimate the magnetic fraction $\eta$ from $B$ and $R$ by dividing total energy injected from magnetar,
\begin{eqnarray}\label{eq:TotalRotationEnergy}
	E_{\rm spin}(t)
	=
	\int^{t}_0 L_{\rm spin}(t') dt'
	\approx
	\frac{2 \pi^2 I}{P^2_0}
	~\mbox{for $P_0 \ll P$},
\end{eqnarray}
by magnetic energy inside MWN,
\begin{eqnarray}\label{eq:TotalBfieldEnergy}
	E_{\rm B}(t)
	& = &
	\frac{B^2 R^3}{6} \left( \frac{t}{\age} \right)^{2 \alpha_{\rm B} + 3 \alpha_{\rm R}}.
\end{eqnarray}
i.e., $\eta \equiv E_{\rm B}(t) / E_{\rm spin}(t)$.
For $P_0 \ll P$ and $(\alpha_{\rm R}, \alpha_{\rm B}) = (1, -1.5)$ (i.e., $2 \alpha_{\rm B} + 3 \alpha_{\rm R} = 0$), $\eta$ is almost constant with time and is compatible with our past studies (TT10, 11, 13b).

From Equations (\ref{eq:ExpansionEvolution}) and (\ref{eq:BfieldEvolution}), the cooling term in Equation (\ref{eq:BoltzmannEquation}) is written as
\begin{eqnarray}\label{eq:CoolingTerms}
	\dot{\gamma}(\gamma, t)
	& = &
	- \frac{\alpha_{\rm R}}{t} \gamma
	- \frac{\gamma^2}{t_{\rm syn}} \left( \frac{t}{\age} \right)^{2 \alpha_{\rm B}}
	- \sum_i \frac{u_i}{u_{\rm B} t_{\rm syn}} \frac{\gamma^2 \gamma^2_{{\rm K}, i}}{\gamma^2 + \gamma^2_{{\rm K}, i}},
\end{eqnarray}
where $i =$ CMB, IR, and OPT and $\gamma_{{\rm K}, i} = 3 \sqrt{5} m_{\rm e} c^2 / 8 \pi k_{\rm B} T_{i}$.
The last term of Equation (\ref{eq:CoolingTerms}) is approximated form of $\dot{\gamma}_{\rm IC}$ given by \citet{Schilickeiser&Ruppel10} and we introduced the current synchrotron time-scale
\begin{eqnarray}\label{eq:SynCoolingTime}
	t_{\rm syn}
	& \equiv &
	\frac{3 m_{\rm e} c}{4 \sigma^{}_{\rm T} u_{\rm B}}
	\approx 2.5 \times 10^8 {\rm kyr}
	\left( \frac{B}{10 \mu {\rm G}} \right)^{-2},
\end{eqnarray}
and $u_{\rm B} \equiv B^2 / 8 \pi$, respectively.
The cooling Lorentz factor $\gamma_{\rm cool} \equiv \alpha_{\rm R} t_{\rm syn} / \age$ will be used to characterize the typical Lorentz factor dividing the dominant cooling process into $\dot{\gamma}_{\rm ad}$ or $\dot{\gamma}_{\rm syn}$ at $t = \age$.


%
\subsection{Particle Injection}\label{sec:ParticleInjection}

Magnetar wind plasma is accelerated and injected into a MWN.
According to past studies of PWNe, we consider that almost all of the spin-down power is converted into the energy of accelerated electron-positron plasma, i.e., the magnetic power is a small fraction of $L_{\rm spin}(t)$ ($\eta \ll 1$).
We adopt a broken power-law injection spectrum of accelerated particles characterized by five parameters $\gamma_{\rm min}, \gamma_{\rm b}, \gamma_{\rm max}, p_1$ and $p_2$ which are the minimum, break, and maximum Lorentz factors and the power-law indices at the low and high energy parts, respectively.
The injection term of Equation (\ref{eq:BoltzmannEquation}) is expressed as
\begin{eqnarray}
	Q_{\rm inj}(\gamma, t)
	& = &
	\chi \frac{L}{\gamma^2_{\rm b} m_{\rm e} c^2}
	\left( \frac{t}{\age} \right)^{-\frac{n+1}{n-1}}
	H(t - t_{\rm s})
	R(\gamma)
	\label{eq:InjectionTerm} \\
	\chi
	& \equiv &
	\gamma^2_{\rm b} \left( \int^{\infty}_1 d \gamma \gamma R(\gamma) \right)^{-1}, \\
	R(\gamma)
	& \equiv &
	\left\{
	\begin{array}{ll}
		(\gamma / \gamma_{\rm b})^{p_1} & \mbox{for $\gamma_{\rm min} \leq \gamma \leq \gamma_{\rm b  }$ ,} \\
		(\gamma / \gamma_{\rm b})^{p_2} & \mbox{for $\gamma_{\rm b  } \leq \gamma \leq \gamma_{\rm max}$ ,}
	\end{array} \right.
	\label{eq:BrokenPL}
\end{eqnarray}
where $H(x)$ is the Heaviside's step function and $\chi$ is a value of order unity for typical sets of parameters $\gamma_{\rm min} \ll \gamma_{\rm b} \ll \gamma_{\rm max}$ and $p_2 < -2 < p_1$.
$Q_{\rm inj}(\gamma, t)$ is normalized to satisfy $L_{\rm spin}(t) = \int d \gamma Q_{\rm inj}(\gamma, t) \gamma m_{\rm e} c^2$.

In Equation (\ref{eq:InjectionTerm}), we introduced an additional parameter, the start time $t_{\rm s} (< \age)$, because we cannot set $t = 0$ in Equations (\ref{eq:CoolingTerms}) and also (\ref{eq:InjectionTerm}).
However, it is possible to set a reasonable finite value of $t_{\rm s}$ from $\age$ and $B$ (Equation (\ref{eq:StartTime})), and we only have to limit our interest to the particle energy range of $\gamma \ge \gamma_{\rm min}$.
As shown in Appendix \ref{app}, the current particle spectrum $N(\gamma \ge \gamma_{\rm min}, \age)$ is not affected by any particles injected before $t = t_{\rm s}$.
On the other hand, we require $t_{\rm s} \gg t_0$ for Equation (\ref{eq:LspinAgeApprox}) to be valid even at $t = t_{\rm s}$.
Combining with Equation (\ref{eq:t0}), the condition $t_{\rm s} > t_0$ constrains the initial period of the magnetar.
From $L \propto \dot{P} P^{-3} \propto t^{-(n+1)/(n-1)}$, we obtain $P \propto t^{1/(n-1)}$ in the same approximation, i.e.,
\begin{eqnarray}\label{eq:ULofP0}
	P(t_{\rm s})
	=
	P \left( \frac{t_{\rm s}}{\age} \right)^{\frac{1}{n-1}}.
\end{eqnarray}
$P(t_{\rm s})$ gives the upper limit on $P_0$ within our formulation.

For typical values of young PWNe, $\gamma_{\rm min}$, $\gamma_{\rm max}$ and $p_1$ have little influence on emissions in radio, X-rays and TeV $\gamma$-rays while $\gamma_{\rm b}$ and $p_2$ remain as the parameters of the injection spectrum.
For the application to 1E 1547.0-5408 in the next section, we adopt $\gamma_{\rm min} = 10^3$, because the characteristic frequencies of synchrotron radiation and inverse Compton scattering off the ISRF $\nu_{\rm IC/ISRF}$ in the Thomson limit are given by \citep[c.f.,][]{Rybicki&Lightman79, Blumenthal&Gould70}
\begin{eqnarray}
	\nu_{\rm syn}
	& \approx &
	1.2 \times 10^7 {\rm Hz}
	\left( \frac{\gamma}{10^3} \right)^2
	\left( \frac{B}{10\mu{\rm G}} \right), \label{eq:SynFreq} \\
	\nu_{\rm IC/ISRF}
	& \approx &
	3.0 \times 10^{20} {\rm Hz}
	\left( \frac{\gamma}{10^3} \right)^2
	\left( \frac{T_{\rm ISRF}}{4000{\rm K}} \right). \label{eq:ICFreq}
\end{eqnarray}
We set $p_1 = -1.5$ which is the typical values to reproduce radio observations of young PWNe (c.f., TT13b).
Lastly, we assume that the maximum energy of the accelerated particles satisfies $\gamma_{\rm max} m_{\rm e} c^2 = e \Phi_{\rm cap}$ \citep[][]{Bucciantini+11}, i.e.,
\begin{eqnarray}\label{eq:MaximumLorentzFactor}
	\gamma_{\rm max}
	& \approx &
	1.3 \times 10^9
	\left( \frac{B_{\rm NS}}{10^{14} {\rm G}} \right)
	\left( \frac{P}{1 {\rm s}} \right)^{-2},
\end{eqnarray}
where the potential difference at the polar cap is estimated as $\Phi_{\rm cap} = \Omega^2 B_{\rm NS} R^3_{\rm NS} / 2 c^2$ \citep[e.g.,][]{Ruderman&Sutherland75}.
Equation (\ref{eq:MaximumLorentzFactor}) essentially corresponds to the Hillas condition $\gamma_{\rm max} m_{\rm e} c^2 = e B(r) r$ \citep[][]{Hillas84} on the assumption that the magnetic field is dipole inside the light cylinder $R_{\rm LC} = c / \Omega$ and is toroidal beyond $R_{\rm LC}$, i.e., the magnetic field strength at an acceleration point $B(r) = B_{\rm NS} (R_{\rm NS} / R_{\rm LC})^3 (R_{\rm LC} / r)$ \citep[c.f.,][]{Goldreich&Julian69}.



\section{Application to 1E 1547.0-5408}\label{sec:1E1547}

1E 1547.0-5408 is an X-ray source discovered in 1980 \citep[][]{Lamb&Markert81} and was recognized as an anomalous X-ray pulsar located inside SNR G327.24-0.13 by \citet{Gelfand&Gaensler07}.
\citet{Camilo+07} discovered radio pulsations of $P = 2.07$ s which is the smallest among magnetars.
For the period derivative, we adopt the long-term average value $\dot{P} \approx 4.77 \times 10^{-11}~{\rm s~s^{-1}}$ \citep[][]{Dib+12, Olausen&Kaspi14}, which is different from $\dot{P}$ obtained by \citet{Camilo+07, Camilo+08}.
The braking index is not well-determined because $\dot{P}$ of 1E 1547.0-5408 shows temporal variations, which is likely associated with flaring activities \citep[][]{Camilo+08, Dib+12}.
Current surface magnetic field strength of $B_{\rm NS} = 3.2 \times 10^{14}~{\rm G}$ is typical while the spin-down power of $L_{\rm spin} = 2.1 \times 10^{35}~{\rm erg~s^{-1}}$ is the largest among magnetars.
The potential difference of the polar cap gives $\gamma_{\rm max} = 1.9 \times 10^9$ (see Equation (\ref{eq:MaximumLorentzFactor})).
We adopt a distance to 1E 1547.0-5408 of $d \sim 4.5~{\rm kpc}$ from the possible association of SNR G327.24-0.13 with a nearby star-forming region \citep[][]{Gelfand&Gaensler07} and from the flux decline of the dust-scattering X-ray rings \citep[][]{Tiengo+10}, although the dispersion measure of $\sim 830~{\rm cm^{-3}~pc}$ \citep[][]{Camilo+07} and the neutral hydrogen column density of $N_{\rm H} \sim 3 \times 10^{22}~{\rm cm^{-2}}$ \citep[][]{Gelfand&Gaensler07, Vink&Bamba09} are relatively large compared with sources of the similar distance.
The characteristic age of $t_{\rm c} = 0.69~{\rm kyr}$ indicates that 1E 1547.0-5408 is one of the youngest pulsars ever observed and is consistent with the small angular size $\sim 4'$ of SNR G327.24-0.13 corresponding to $\sim 5.2~{\rm pc}$ in diameter \citep[][]{Gelfand&Gaensler07}.

\citet{Vink&Bamba09} reported the discovery of extended X-ray emission around 1E 1547.0-5408 from archival {\it Chandra} and {\it XMM-Newton} data in 2006 with the source in quiescence.
They interpreted the extended emission as a PWN whose angular radius of $\sim 45''$ ($\sim 0.98~{\rm pc}$ in radius) and the 2--10 keV flux of $F_{\rm VB09} = (1.5 \pm 0.3) \times 10^{-13}~{\rm erg~s^{-1}~cm^{-2}}$.
However, both its soft spectrum (photon index of $\Gamma_{\rm X} \sim 3.5$) and a large X-ray efficiency of $\eta_{\rm X} = (4 \pi d^2 F_{\rm VB09}) / L_{\rm spin} \sim 1.7 \times 10^{-3}$ are not typical as young PWNe.
\citet{Olausen+11} reanalyzed the {\it XMM-Newton} observation in 2006 together with 2007, 2009 and 2010 data.
They also found extended emission but its flux is different between observations.
As already reported by \citet{Tiengo+10} for the 2009 event, the extended emission on the 2007, 2009 and 2010 data was interpreted as the dust-scattering halo.
For the 2006 data, \citet{Olausen+11} did not rule out the presence of a faint PWN and gave the stronger upper limit on the 2--10 keV flux $\lesssim 4.7 \times 10^{-14}~{\rm erg~s^{-1}~cm^{-2}} \equiv F_{\rm O11}$ than $F_{\rm VB09}$.
Although 1E 1547.0-5408 has been extensively observed in radio, infrared, hard X-rays and GeV $\gamma$-rays \citep[c.f.,][]{Olausen&Kaspi14}, we do not find other flux information of its extended emission.

We summarize the parameters of the calculations (Figures \ref{fig:model1} and \ref{fig:model2}) together with Kes 75 parameters obtained by TT11 in Table \ref{tbl:Parameters}.
Since Kes 75 shows some exceptional features among young PWNe around RPPs, we expect that the wind from high B-field pulsar (HBP) PSR J1846-0258 is similar to that of magnetars rather than RPPs \citep[c.f.,][]{Safi-Harb13}.
$\gamma_{\rm min}$ and $p_1$ are fixed parameters and then we do not obtain information of the pair multiplicity of the wind \citep[c.f., TT10;][]{deJager07, Bucciantini+11}.
The dependent parameters, such as the ratio $\gamma_{\rm cool} / \gamma_{\rm max}$, the age of the system $\age$, the start time $t_{\rm s}$, and the upper limits on the initial period $P_0$ (Equation (\ref{eq:ULofP0})) and on the magnetic energy fraction $\eta$ (Equations (\ref{eq:TotalRotationEnergy}) and (\ref{eq:TotalBfieldEnergy})), are also tabulated.

\subsection{Results}\label{sec:1E1547result}

%
\begin{figure}
\includegraphics[scale=0.6]{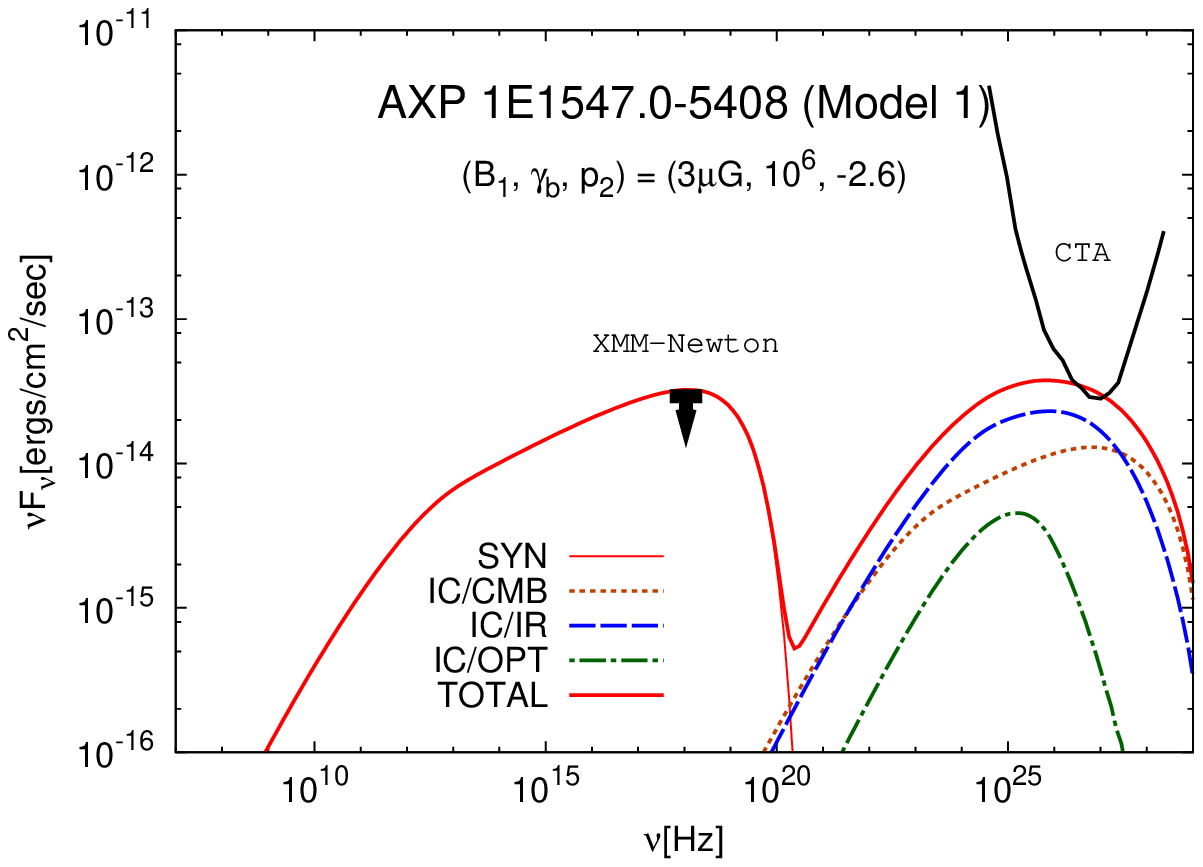}
\includegraphics[scale=0.6]{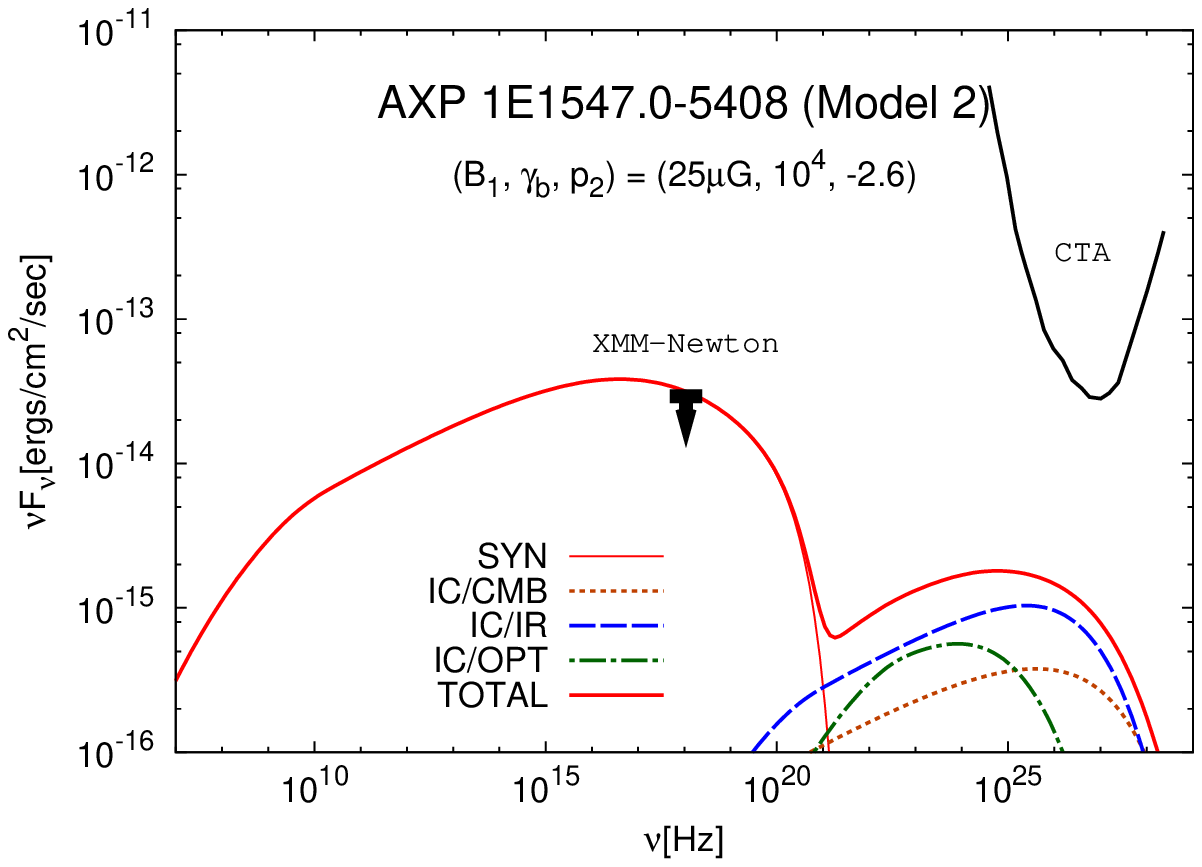}
\caption{
	Model spectra (Models 1 and 2) of the MWN surrounding AXP 1E 1547.0-5408 with the upper limit in 2 -- 10 keV given by {\it XMM-Newton} \citep[][]{Olausen+11} and the sensitivity of {\it CTA} \citep[50h,][]{Acharya+13}.
	The parameters of the calculations are tabulated in Table \ref{tbl:Parameters} for Model 1 (left panel) and Model 2 (right panel), respectively.
	The thick red line is the total spectra which is the sum of the synchrotron (thin red), IC/CMB (dotted orange), IC/IR (dashed blue) and IC/OPT (dot-dashed green) components.
\label{fig:model1}
}
\end{figure}
\begin{figure}
\includegraphics[scale=0.6]{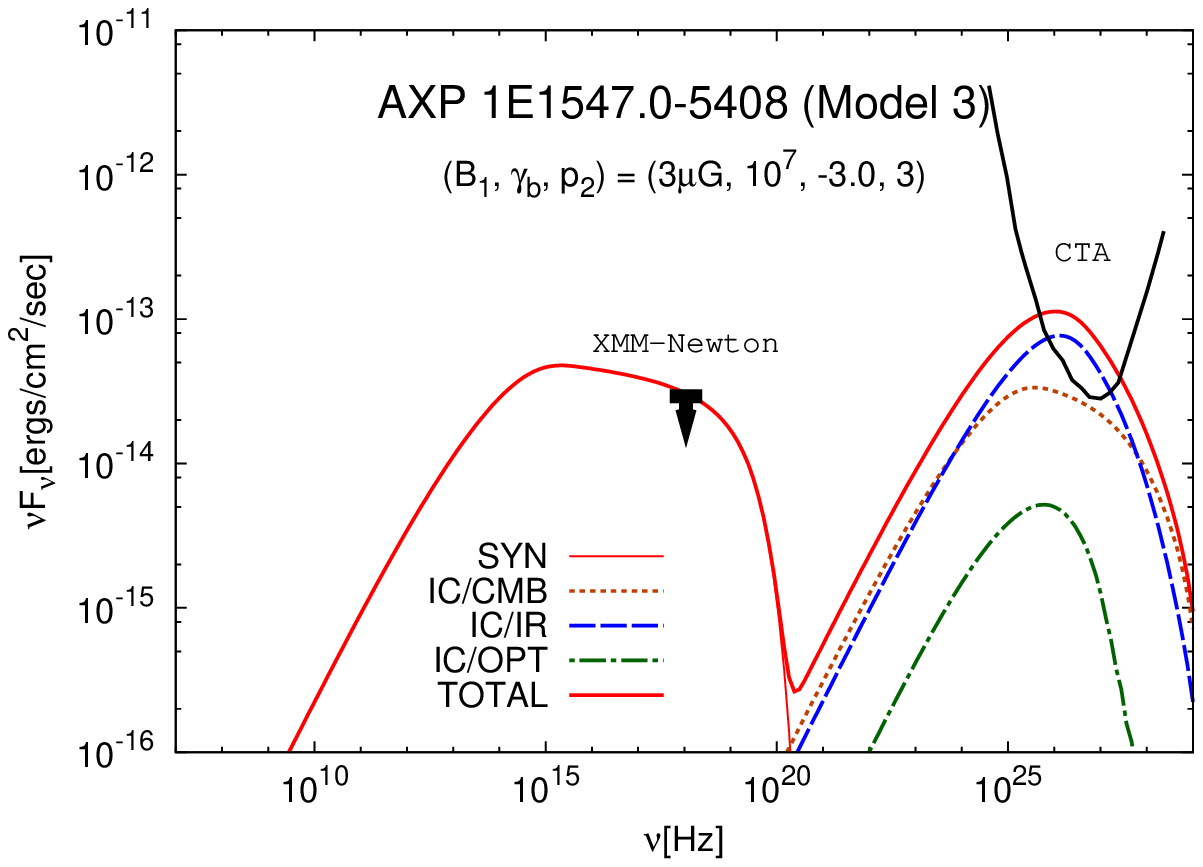}
\includegraphics[scale=0.6]{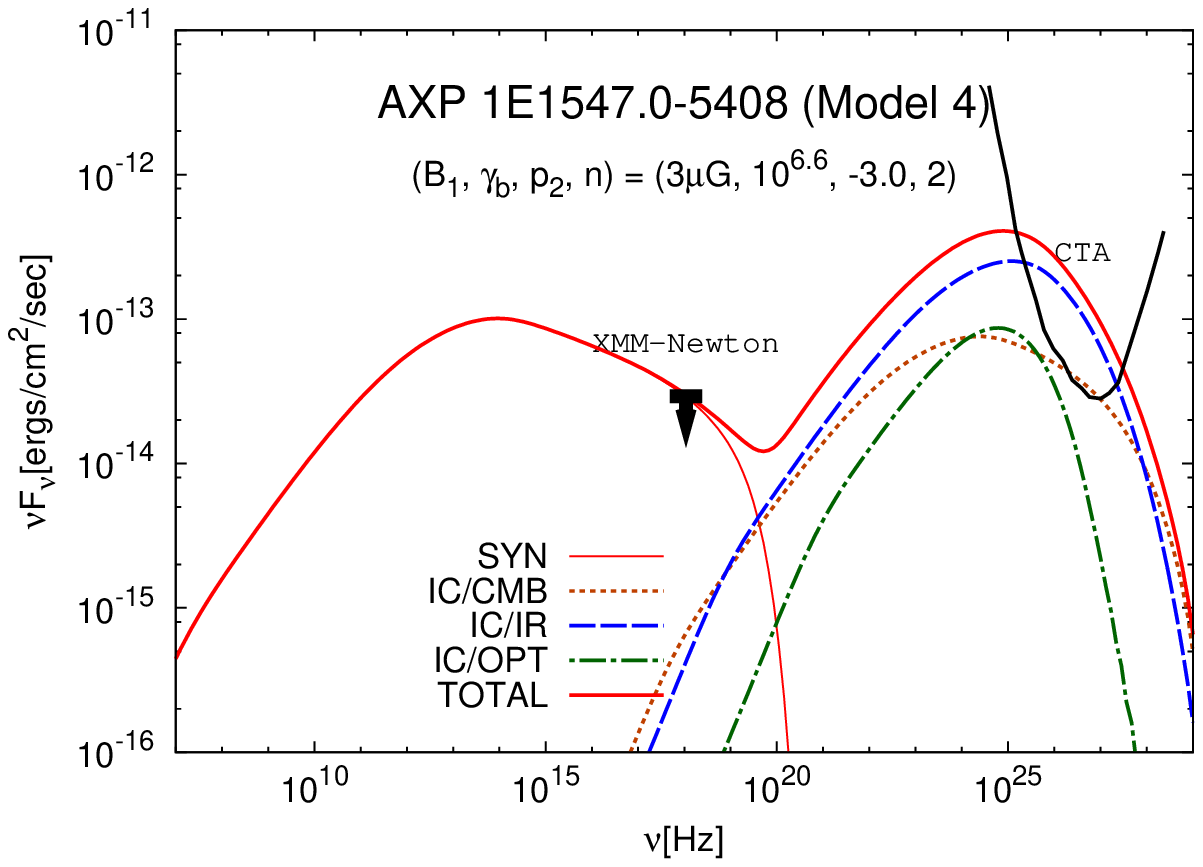}
\caption{
	Model spectra (Models 3 and 4) of the MWN surrounding AXP 1E 1547.0-5408.
	The parameters of the calculations are tabulated in Table \ref{tbl:Parameters}.
\label{fig:model2}
}
\end{figure}
\begin{table}[!t]
\caption{
	Summary of the parameters for the calculations in Figures \ref{fig:model1} and \ref{fig:model2}.
	The parameters of Kes 75 is taken from TT11.
}
\label{tbl:Parameters}
\begin{tabular}{crrrrr}
	Symbol  &
	Kes 75  &
	Model 1 &
	Model 2 &
	Model 3 &
	Model 4 \\
\hline
\multicolumn{6}{c}{Fixed Parameters} \\
\hline
$d$(kpc)                        & 6.0     & \multicolumn{4}{c}{4.5} \\
$R$(pc)                         & 0.29    & \multicolumn{4}{c}{0.98} \\
$P$(s)                          & 0.326   & \multicolumn{4}{c}{2.07} \\
$\dot{P}$($10^{-12}$s s$^{-1}$) & 7.08    & \multicolumn{4}{c}{47.7} \\
$u_{\rm IR}$(eV cm$^{-3}$)      & 1.2     & \multicolumn{4}{c}{1.0} \\
$u_{\rm OPT}$(eV cm$^{-3}$)     & 2.0     & \multicolumn{4}{c}{2.0} \\
$\gamma_{\rm min}(10^3)$        & $<$ 5.0 & \multicolumn{4}{c}{1.0} \\
$\gamma_{\rm max}(10^9)$        & $>$ 1.0 & \multicolumn{4}{c}{1.9} \\
$p_1$                           & -1.6    & \multicolumn{4}{c}{-1.5} \\
$\alpha_{\rm R}$                & 1.0     & \multicolumn{4}{c}{1.0} \\
$\alpha_{\rm B}$                & -1.5    & \multicolumn{4}{c}{-1.5} \\

\hline
\multicolumn{6}{c}{Adopted Parameters} \\
\hline
$B(\mu {\rm G})$     & 20   & 3    & 25   & 3    & 3    \\
$\gamma_{\rm b}(10^5)$ & 20   & 10   & 0.1  & 100  & 60   \\
$p_2$                  & -2.5 & -2.6 & -2.6 & -3.0 & -3.0 \\
$n$                    & 2.65 & 3    & 3    & 3    & 2    \\

\hline
\multicolumn{6}{c}{Dependent Parameters} \\
\hline
$\gamma_{\rm cool}/\gamma_{\rm max}$ &      & 2.1      & 0.030   & 2.1      & 1.0         \\
$\age$(kyr)                       & 0.7  & 0.69     & 0.69    & 0.69     & 1.4         \\
$t_{\rm s}$(yr)                   &      & 3.0      & 12      & 3.0      & 7.6         \\
$P_0$(ms)                         & 126  & $<$ 137  & $<$ 278 & $<$ 137  & $<$ 11      \\
$\eta(10^{-3})$                   & 0.05 & $<$ 0.04 & $<$ 11  & $<$ 0.04 & $<$ 0.00028 \\

\end{tabular}

\end{table}

Figure \ref{fig:model1} shows model spectra of the MWN surrounding 1E 1547.0-5408 with the 2--10 keV flux upper limit $F_{\rm O11}$ given by \citet{Olausen+11}.
We fix $(p_2, n) = (-2.6, 3)$ and then we search for the combinations of $(B, \gamma_{\rm b})$ for the calculated X-ray flux to be close to $F_{\rm O11}$.
The left panel of Figure \ref{fig:model1} (Model 1) is the case of $(B, \gamma_{\rm b}) = (3 \mu {\rm G}, 10^6)$, where $\gamma_{\rm b}$ is close to that of Kes 75.
We require the weak magnetic field strength comparable to the interstellar magnetic field $\approx 3 \mu {\rm G}$ to suppress the X-ray flux below the upper limit $F_{\rm O11}$.
The larger magnetic field is allowed for the smaller $\gamma_{\rm b}$.
The right panel of Figure \ref{fig:model1} (Model 2) is the case of $(B, \gamma_{\rm b}) = (25 \mu {\rm G}, 10^4)$, where the value of $B$ is close to that of Kes 75.
Since $\gamma_{\rm b} = 10^4$ is the smallest among young PWNe ever studied (c.f., TT13b), the mean magnetic field strength of less than $25 \mu {\rm G}$ is favored for $(p_2, n) = (-2.6, 3)$.

The two spectra in Figure \ref{fig:model1} are different in radio and $\gamma$-ray bands.
Model 2 predicts extended ($\sim 45''$ in radius) radio nebula of the total flux $\sim$ 300 mJy at 1 GHz and $\sim$ 60 mJy at 10 GHz, while the total radio flux is about an order of magnitude smaller for Model 1.
Note that the radio flux depends on $(p_1, \gamma_{\rm min})$ which we fixed in this paper.
On the other hand, Model 1 is far brighter than Model 2 in TeV $\gamma$-rays and will be detected by {\it CTA}.
Future deep observations in X-rays would also distinguish models because the photon indices are different between models.
Model 1 has the harder spectrum than Model 2 because $\gamma_{\rm cool} / \gamma_{\rm max} \sim 1$ for Model 1, i.e., a synchrotron cooling break signature does not appear in the spectrum.

Since some young PWNe have the soft injection spectrum of $p_2 \sim -3.0$, we also study such cases.
Figure \ref{fig:model2} shows model spectra for the small magnetic field cases $(B, p_2) = (3 \mu {\rm G}, -3.0)$.
The braking index is different for the left ($n = 3$) and right ($n = 2$) panels and then we search for the value of $\gamma_{\rm b}$ for the calculated X-ray flux to be close to $F_{\rm O11}$.
In both cases, the large TeV $\gamma$-ray flux is predicted compared with Figure \ref{fig:model1} without increasing the energy density of the ISRF $(u_{\rm IR}, u_{\rm OPT})$.

\section{DISCUSSION AND CONCLUSIONS}\label{sec:dis&cons}

Considering the plasma outflows from the magnetar, we built a broadband emission model of MWNe based on the one-zone model of young PWNe around RPPs (TT10, TT11, TT13b).
The model is simplified for the application to MWNe that have less observational information than PWNe.
We apply the model to the MWN around 1E 1547.0-5408 that is the most promising object to detect MWNe among all the known magnetars because of its large $L_{\rm spin}$ and its small distance to the Earth.

Because of poor current observational constraints on the MWN around 1E 1547.0-5408, various combinations of parameters are allowed.
However, here, we compare Models 1 and 2 based on the past studies of young PWNe, especially focusing on Kes 75 around PSR J1846-0258 that is an only HBP showing a magnetar-like behavior \citep[][]{Gavriil+08}.
Model 1 has similar $\eta$ and $\gamma_{\rm b}$ with Kes 75, while Model 2 has similar magnetic field strength $B$.
It is known that the mean magnetic field $B$ is very different for each young PWN $3 \lesssim B \lesssim 80 \mu {\rm G}$ \citep[c.f., TT13b;][]{Torres+14}.
On the other hand, TT11 and TT13b showed that $\eta \sim 10^{-3}$ is common between young PWNe except for Kes 75 by using almost the same model of this study, i.e., adopting $(\alpha_{\rm R}, \alpha_{\rm B}) = (1.0, -1.5)$ in Equations (\ref{eq:ExpansionEvolution}), (\ref{eq:BfieldEvolution}) and (\ref{eq:TotalBfieldEnergy}).
Kes 75 has exceptionally small $\eta$ compared with other young PWNe for our model and also other models, although the values of $\eta$ are different between models \citep[][]{Bucciantini+11, Torres+14}.
We favor Model 1 over Model 2 because we expect that magnetars have the similar wind property $\eta$ and $\gamma_{\rm b}$ to HBPs, where $\gamma_{\rm b}$ is related with the bulk Lorentz factor of the wind just upstream the termination shock \citep[e.g.,][]{Kennel&Coroniti84, Bucciantini+11}.

Models 3 and 4 are the cases for the soft injection spectrum $p_2 = -3$ and the large break Lorentz factor $(\gamma_{\rm b} \gg 10^6)$ although such $\gamma_{\rm b}$ and $p_2$ are not typical among young PWNe (c.f., TT13b).
Both models are interesting because they predict the larger TeV $\gamma$-ray flux than Model 1 without increasing the local ISRF $(u_{\rm IR}, u_{\rm OPT})$.
For example, the $\gamma$-ray flux increases, if the photons from the nearby star-forming region contribute to the local ISRF \citep[c.f.][]{Torres+14}.
Model 3 is brighter than Model 2 in TeV $\gamma$-rays because the injection of the high energy particles $Q_{\rm inj}(\gamma > \gamma_{\rm b}) \propto \gamma^{-p_2 - 2}_{\rm b}$ is an increasing function of $\gamma_{\rm b}$.
Model 4 predicts larger TeV $\gamma$-ray flux than Model 3 because the braking index changes $P(t_{\rm s})$ (Equation (\ref{eq:ULofP0})), i.e., the larger amount of the rotational energy is injected for Model 4 than Model 3.
The models will be distinguished by future observations by {\it CTA}.
Note that we postulate $B = 3\mu {\rm G}$ and the X-ray flux to be the level of $F_{\rm O11}$ for Models 3 and 4.
On the other hand, the soft spectrum $p_2 = -3$ with $B \gg 3 \mu{\rm G}$ and $10^4 \lesssim \gamma_{\rm b} \ll 10^6$ predicts dark MWNe that are difficult to detect with current or near future instruments in both X-rays and TeV $\gamma$-rays.
In other words, a non-detection of the MWN around 1E 1547.0-5408 does not simply reject the existence of magnetar winds.

Contrary to the studies of young PWNe, the current model gives only upper limits on the initial period of magnetar $P_0$ \citep[c.f.,][]{deJager08}.
The upper limits $P(t_{\rm s})$ for each model are tabulated on Table \ref{tbl:Parameters}.
From population synthesis studies of isolated neutron stars \citep[e.g.,][]{Faucher-Giguere&Kaspi06, Popov+10, Igoshev&Popov13} and from some estimated initial periods of individual pulsars \citep[e.g., TT11; TT13b;][]{Popov&Turolla12}, the condition $P_0 < 100$ ms is usually satisfied for most of pulsars.
We consider that the simplified spin-down (Equation (\ref{eq:LspinAgeApprox})) is the reasonable approximation for MWNe.
This is different from the cases of young PWNe whose initial spin-down time $t_0$ is mostly close to an age of the systems (e.g., TT11; TT13b).
Nevertheless, if $P_0 > P(t_{\rm s})$, we should use Equation (\ref{eq:Lspin}).
The use of Equation (\ref{eq:Lspin}) simply reduces the total injection of the rotational energy compared with Equation (\ref{eq:LspinAgeApprox}) and the MWN becomes dark.
This is another possible reason for a non-detection of the MWN around 1E 1547.0-5408.

Our simplified model of a MWN spectrum does not allow us to test the existence of millisecond magnetars because $P(t_{\rm s})$ is much larger than milliseconds for 1E 1547.0-5408.
Severe adiabatic and synchrotron cooling effects dismiss the non-thermal emission from the particles injected at $t < t_{\rm s}$ unless considering a re-acceleration of these particles, for example.
However, the magnetic field inside the MWN is found from the combination of $P_0$ and $\eta$.
From Equations (\ref{eq:TotalRotationEnergy}) and (\ref{eq:TotalBfieldEnergy}) with $(\alpha_{\rm R}, \alpha_{\rm B}) = (1, -1.5)$, we obtain
\begin{eqnarray}
	B_{\rm MWN}
	& \approx &
	210~\mu {\rm G}
	\left( \frac{\eta       }{10^{-5}      } \right)^{ \frac{1}{2}}
	\left( \frac{R_{\rm MWN}}{0.98~{\rm pc}} \right)^{-\frac{3}{2}}
	\left( \frac{P_0        }{1~{\rm ms} }   \right)^{-1          }.
\end{eqnarray}
Adopting $\eta \sim 10^{-5}$ for the wind from magnetars and HBPs, the mean magnetic field $\approx 200 \mu {\rm G}$ for $P_0 \approx$ 1 ms does not conflict with the upper limit in X-rays when the parameters of the injection spectrum are $p_2 = -3$ and $\gamma_{\rm b} = 10^4$, for example.
Note that this estimate of $B_{\rm MWN}$ is strongly depends on the choice of $(\alpha_{\rm R}, \alpha_{\rm B})$ and this is the reason why we do not use $\eta$ as a parameter in the present model.

Lastly, we discuss particle injections accompanied with flaring activities of magnetars.
The transient radio nebula was detected following the giant flares of SGR1806-20 \citep[][]{Cameron+05, Gaensler+05} and also of SGR1900+40 \citep[][]{Frail+99}.
Although the isotropic energy of the strongest giant flare from SGR1806-20 attains to $\sim 10^{47} {\rm erg}$ \citep{Palmer+05, Hurley+05, Terasawa+05}, it is an exceptionally large event.
Considering that a typical giant flare has the isotropic energy ranging $\sim 10^{45 - 47}$ erg and has the event rate of once per 50$-$100 yr per source \citep[c.f.,][]{Woods&Thompson06}, the total energy associated with giant flares is $\lesssim 10^{48}$ erg for $\age \sim$ kyr.
On the other hand, the total rotational energy injected from $t_{\rm s}$ attains to $2 \pi^2 I / P^2(t_{\rm s}) \approx 10^{48} {\rm erg}$ for Model 1.
We consider that the particle energy injection associated with giant flares do not dominate over persistent particle injection by the magnetar wind for 1E 1547.0-5408.

Magnetars also show the bursting activities less luminous than the giant flares but they are much more frequent.
Although the cumulative energy distributions of magnetar bursts are a power-law distribution $d N / d E \propto E^{-\beta}$ with $\beta < 2$ \citep[][]{Cheng+96, Gogus+99, Gogus+00}, it is argued that the persistent X-ray emission of magnetars $L_{\rm X} (> L_{\rm spin})$ might be the accumulation of unresolved small bursts \citep[e.g.,][]{Enoto+12}.
If there are the particle injections associated with these short bursts, observed $L_{\rm X}$ larger than $L_{\rm spin}$ may be interpreted as the wind loss model \citep[]{Harding+99, Thompson+00, Thompson+02}.
In their models, the wind luminosity $L_{\rm wind}$ is allowed to be significantly larger than the power estimated from observed period and its derivative $I \Omega \dot{\Omega} = L_{\rm spin}$.
Note that, because $L_{\rm X}$ of 1E 1547.0-5408 in quiescence phase is less than $L_{\rm spin}$, an episodic particle wind model by \citet{Harding+99} should be applied.
We simply expect that the particle luminosity and also the radiation from the MWN are proportional to $L_{\rm wind}$ and then we require $B \lesssim 3 \mu{\rm G}$, $p_2 < -2.6$ and/or $\gamma_{\rm b} < 10^6$ for the calculated X-ray flux to be below the upper limit.
In addition, the large $\gamma$-ray flux is also expected in this case.

\section*{Acknowledgments}

S. J. T. would like to thank S. Kisaka, T. Enoto, K. Asano, and T. Terasawa for useful discussion.
S. J. T. would also like to thank anonymous referee for his/her very helpful comments.
This work is supported by JSPS Research Fellowships for Young Scientists (S. T., 2510447).

\bibliography{draft}

\appendix

\section{Appendix: Reduction of Model Parameters}\label{app}

\begin{figure}
\includegraphics[scale=1.0]{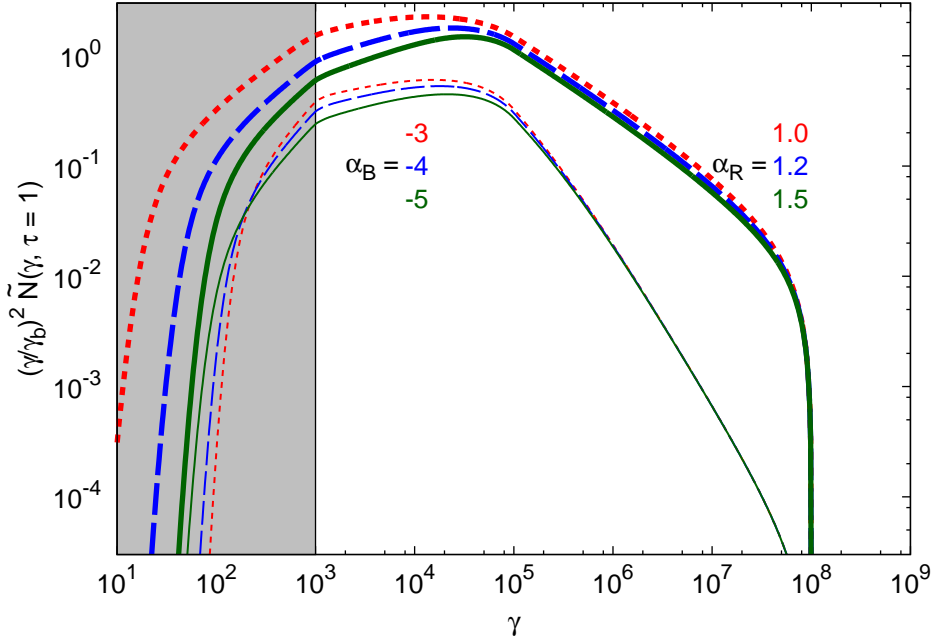}
\caption{
	Particle energy distributions $(\gamma/\gamma_{\rm b})^2 \tn(\gamma, \tau = 1)$ calculated from Equation (\ref{eq:NormalizedBoltzmannEquation}).
	The parameters of the injection spectrum $(\gamma_{\rm min}, \gamma_{\rm b}, \gamma_{\rm max}, p_1, p_2) = (10^3, 10^5, 10^8, -1.5, -2.5)$ are common for all the lines and $n = 3$ is adopted.
	We are not interested in the shaded region corresponding to the particle energy of $\gamma < \gamma_{\rm min}$ because we take $\tau_{\rm s}$ according to Equation (\ref{eq:StartTime}).
	For three upper thick lines, $\alpha_{\rm B} = -1.5$ and $\tau_{\rm syn} = 10^8$ are common and $\alpha_{\rm R}$ is different 1.0 (red dotted), 1.2 (blue dashed) and 1.5 (green solid) for each line.
	All the three spectra are similar to each other especially in high energy range which we are interested in this paper.
	For three lower thin lines, $\alpha_{\rm R} = 1.0$ and $\tau_{\rm syn} = 10^5$ are common and $\alpha_{\rm B}$ is different -1.5 (red dotted), -2.0 (blue dashed) and -2.5 (green solid) for each line.
	All the three spectra are also similar to each other.
\label{fig:AlphaBRDependence}
}
\end{figure}

The model presented in Section \ref{sec:Model} has many parameters.
Here, we show that $\alpha_{\rm R}, \alpha_{\rm B}$ and $t_{\rm s}$ can be chosen not to affect the calculated current spectra.
For simplicity, we ignore $\dot{\gamma}_{\rm IC}$ in Equation (\ref{eq:CoolingTerms}) because $\dot{\gamma}_{\rm IC}$ does not the dominant cooling process.
It is convenient to measure the times $t, t_{\rm s}$, and $t_{\rm syn}$ with $\age$ ($\tau, \tau_{\rm s}$, and $\tau_{\rm syn}$) and the particle number $N(\gamma, t)$ with $\chi L \age / \gamma^2_{\rm b} m_{\rm e} c^2$ ($\tn(\gamma, \tau)$).
These normalization are unique for a given central magnetar $(\age, L)$ and for a given set of $(\gamma_{\rm min}, \gamma_{\rm b}, \gamma_{\rm max}, p_1, p_2)$.
Equation (\ref{eq:BoltzmannEquation}) becomes
\begin{eqnarray}
	\frac{\partial}{\partial \tau} \tn(\gamma, \tau)
	+
	\frac{\partial}{\partial \gamma} \left[ \left( \frac{d \gamma}{d \tau} \right) \tn(\gamma, \tau) \right]
	& = &
	\tau^{-\frac{n+1}{n-1}} H(\tau - \tau_{\rm s})
	R(\gamma),
	\label{eq:NormalizedBoltzmannEquation} \\
	\frac{d \gamma}{d \tau}
	& \equiv &
	- \frac{\alpha_{\rm R}}{\tau} \gamma - \frac{\tau^{2 \alpha_{\rm B}}}{\tau_{\rm syn}} \gamma^2.
	\label{eq:NormalizedCoolingTerm}
\end{eqnarray}
The normalized Equation (\ref{eq:NormalizedBoltzmannEquation}) has only four parameters $\tau_{\rm s}, \tau_{\rm syn}, \alpha_{\rm R}$ and $\alpha_{\rm B}$.
One significant parameter is $\tau_{\rm syn}$ which corresponds to the cooling Lorentz factor $\gamma_{\rm cool}$ determined by the magnetic field strength $B$.
Below, we study the role of the rest of three parameters.

Limiting our interest to the particles whose Lorentz factor is larger than $\gamma_{\rm min}$ at $\tau = 1$, the start time $\tau_{\rm s}$ is determined as follows.
Here, we consider the cooling evolution of an accelerated particle whose Lorentz factor is $\gamma_{\rm s}$ at a time $\tau_{\rm s}$.
Equation (\ref{eq:NormalizedCoolingTerm}) represents the cooling evolution and has the analytic solution \citep[c.f.,][]{Pacini&Salvati73}
\begin{eqnarray}\label{eq:ParticleEnergyEvolution}
	\gamma(\tau)
	& = &
	\gamma_{\rm s}
	\left( \frac{\tau_{\rm s}}{\tau} \right)^{\alpha_{\rm R}}
	\left(
		\frac{\gamma_{\rm s} \tau^{\alpha_{\rm R}}_{\rm s}}{\xi \tau_{\rm syn}}
		\left(
		\tau^{-\xi}_{\rm s}
		-
		\tau^{-\xi}
	\right)
	+ 1
	\right)^{-1} \nonumber \\
	& \approx &
	\xi \tau_{\rm syn}
	\tau^{-\alpha_{\rm R}}
	(\tau^{-\xi}_{\rm s} - \tau^{-\xi})^{-1}
	~\mbox{for $\gamma_{\rm s} \rightarrow \infty$},
\end{eqnarray}
where $\xi \equiv \alpha_{\rm R} - 2 \alpha_{\rm B} - 1  > 0$, for example, $\xi =$ 3 and 5.5 for $(\alpha_{\rm R}, \alpha_{\rm B}) =$ (1.0, -1.5) and (1.5, -2.5), respectively.
From the last expression in Equation (\ref{eq:ParticleEnergyEvolution}), we find that all the particles injected at $\tau = \tau_{\rm s}$ have the Lorentz factor less than $\gamma_{\infty}(\tau_{\rm s}) \equiv \xi \tau_{\rm syn} / (\tau^{-\xi}_{\rm s} - 1)$ at $\tau = 1$.
For our purpose, we obtain the start time $\tau_{\rm s}$ by solving $\gamma_{\infty}(\tau_{\rm s}) = \gamma_{\rm min}$, i.e.,
%
\begin{eqnarray}\label{eq:StartTime}
	\tau_{\rm s}
	& = &
	\left(
		1 + \frac{\xi \tau_{\rm syn}}{\gamma_{\rm min}}
	\right)^{-\frac{1}{\xi}}
	\approx
	\left(
		\frac{\gamma_{\rm min}}{\xi \tau_{\rm syn}}
	\right)^{\frac{1}{\xi}} ~ \mbox{for $\tau_{\rm s} \ll 1$,}
	\\
	& \approx &
	\left\{
	\begin{array}{ll}
		\displaystyle{
			1.5 \times 10^{-2}
			(\tau_{\rm syn}/10^8)^{-1/3}
			(\gamma_{\rm min}/10^3)^{1/3}
		}
	& \mbox{for $\xi = 3.0$,} \\
		\displaystyle{
			9.0 \times 10^{-2}
			(\tau_{\rm syn}/10^8)^{-2/11}
			(\gamma_{\rm min}/10^3)^{2/11}
		}
	& \mbox{for $\xi = 5.5$.} \\
	\end{array} \right. \nonumber
\end{eqnarray}
%
From the last expressions, $\tau_{\rm s}$ weakly depends on $\alpha_{\rm R}, \alpha_{\rm B}, \tau_{\rm syn}$ and also $\gamma_{\rm min}$.
Adopting $\tau_{\rm s}$ given by Equation (\ref{eq:StartTime}), the signatures of the particles injected before $\tau_{\rm s}$ does not appear in $\tn(\gamma, \tau)$ at $\tau = 1$ in the range of $\gamma \ge \gamma_{\rm min}$.

The particle spectrum $\tn(\gamma, \tau)$ at $\tau = 1$ is not sensitive to $(\alpha_{\rm R}, \alpha_{\rm B})$ for given $\tau_{\rm syn}$.
In Figure \ref{fig:AlphaBRDependence}, we study dependence of $\tn(\gamma, 1)$ on $\alpha_{\rm R}$ and $\alpha_{\rm B}$, where we set $(\gamma_{\rm min}, \gamma_{\rm b}, \gamma_{\rm max}, p_1, p_2) = (10^3, 10^5, 10^8, -1.5, -2.5)$ and $n = 3$.
Upper thick lines are $(\alpha_{\rm B}, \tau_{\rm syn}) = (-1.5, 10^8)$ and study dependence on $\alpha_{\rm R}$ for three different values 1.0 (dotted), 1.2 (dashed) and 1.5 (solid).
Lower thin lines are $(\alpha_{\rm R}, \tau_{\rm syn}) = (1.0, 10^5)$ and study dependence on $\alpha_{\rm B}$ for three different values -1.5 (dotted), -2.0 (dashed) and -2.5 (solid).
We find no significant difference of $\tn(\gamma, \tau = 1)$ for different values of both $\alpha_{\rm R}$ and $\alpha_{\rm B}$, respectively.
Note that particles of $\gamma \gtrsim 10^5$ are responsible for the X-ray and TeV $\gamma$-ray emission.
We adopt $(\alpha_{\rm R}, \alpha_{\rm B}) = (1.0, -1.5)$ in this paper.

\end{document}